# >Negatively Charged Silver-Doped Gold Nanoscale Clusters

Vitaly Chaban[1]

1) Instituto de Ciência e Tecnologia, Universidade Federal de São Paulo, 12231-280, São José dos Campos, SP, Brazil

2) Department of Chemistry, University of Southern California, Los Angeles, CA 90089, United States

**Abstract**. Gold nanoparticles (GNPs) constitute a breakthrough in modern chemistry. The recent success in the synthesis and total structure determination of the precise-composition GNPs provides exciting opportunities for fundamental studies and development of novel applications. GNPs allow for specific number of alien metal atoms to be incorporated into the gold core. Doping tunes optical, structural, and electronic properties of the nanostructure. This work examines silver-doped negatively charged $Au_{25}Ag_{25-N}$ (N=1,2,3) GNPs using PM7-MD. GNP is shown to accommodate multiple silver atoms, whereas silver atoms prefer to create $Au_2$ clusters at the surface. Silver-gold binding appears inferior to the gold-gold binding.

**Key words**: gold, nanostructure, doping, silver, structure, band gap, semiempirical, PM7-MD.

---

[1] E-mail: vvchaban@gmail.com

**Research Highlights**

1. Structures of the silver-doped gold nanoparticles are computed from PM7-MD simulations.

2. The derived pair correlation functions are helpful to interpret experimental data.

3. Silver atoms prefer to form clusters at the surface rather than coordinating gold atoms.

TOC Image

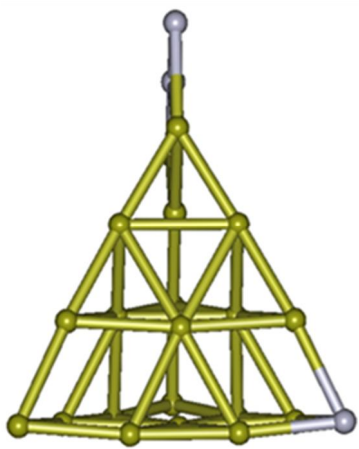 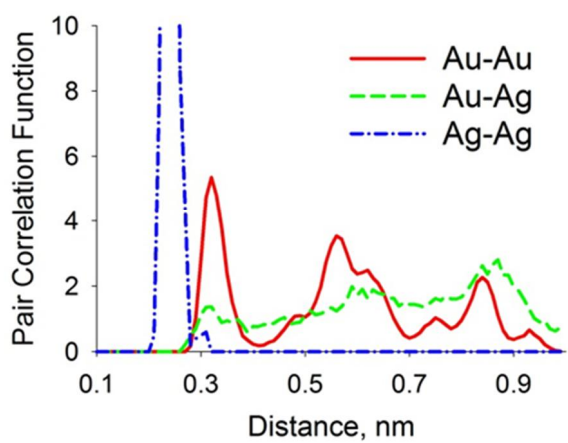

**Introduction**

Gold is omnipresent in our conscience. Recently gold nanoscale particles (GNPs) emerged[1-12] and found their niche within biomedical applications, including genomics, biosensorics, immunoassays, laser phototherapy of cancer cells and tumors, targeted delivery of drugs, optical bio-imaging and so on.[13-15] Such abundant already existing and prospective applications are, first of all, due to unique optical properties of the quantum-sized nanoclusters. GNPs exhibit specific atom-packing, discrete electronic structure, single-electron transitions in optical absorption and, furthermore, a remarkable catalytic behavior.[1,2,15]

Nevertheless, a lot of fundamental issues still persist, which cannot be directly addressed. It is basically unknown why certain types/compositions of nanoparticles are particularly stable as compared to others, if one goes beyond consideration of magic numbers. What exactly protects nanoparticles? How are the particles protected by ligands and surfactants? How can this chemical protection be modulated in different environments? None of these global questions currently have clear answers.

Transmission electron microscopy reveals particle morphology and sometimes shape. It permits quite precise measurements of size (±0.2 nm). Electron microscopy is, however, totally forceless in the determination of nanoparticle surface, e.g. which atoms prevail at the surface and how they are bonded to the core. In turn, spectroscopic techniques offer useful information about ligands. This information alone is always somewhat ambiguous, since nanoparticles coexist in a mixture with excess surfactants and ligands. Other impurities and by-products are often present. They cannot be easily separated. Theoretical simulations can contribute heavily in this situation thanks to their principal ability to isolate effects and offer a truly atomistic precision in addition to femtosecond time resolution of every simulated process.

The achieved success of the experimental techniques to obtain highly uniform GNPs must be acknowledged. Indeed, the best cases were proven to offer size distributions of just five percent. Unfortunately, determination of brutto formulas of these highly uniform GNPs remains extremely challenging. It is often the case when studies employing electrospray ionization mass spectrometry provide somewhat different compositions of the same GNP. Since many GNPs are not neutral, i.e. coexist in solution with the counter-ion, the overall charge of the cluster is of key importance. It is critical to determine an accurate formula of the core and number of ligands to proceed with analysis.

In addition, determination of the doped nanoparticle structures is highly challenging. Even with brutto chemical formula being determined, location of the dopant atoms (at the center? at the surface?) is unknown. Furthermore, information about chemical bonding and its strength is missed. Does Ag-Ag bond exist? How strong is Ag add-atom binding to the rest of GNP? How significant is the effect of a single add-atom (on electronic structure, etc) and how it modulates with the dopant content increase? How many dopants can coexist without breaking GNP? These problems can be solved using the PM7-MD simulations.[16-19] Molecular dynamics (MD) is a necessary constituent of the methodology, since all geometry optimization algorithms are able only to locate the closest local minimum. Nanostructures are extremely rich in local minima, which must be efficiently scanned to locate the deepest and most realistic ones. Sampling all or, at least, most of them is required to obtain a realistic picture of the nanostructure. Such sampling can be achieved using MD at finite temperature.

PM7-MD was successfully applied before to investigate ion solvation,[17] carbon dioxide capture,[19] gold nanocluster coverage[18] and global minimum search.[16] Its application in the present work in favor of phenomenological interaction potentials is justified by complexity of parameterization of such potentials for d-metals.

**Methodology**

Six systems (precise atomistic compositions) and thirty two independent PM7-MD simulations[16-19] of the silver-doped GNPs are discussed in this work (Table 1). $Au_{25}$ GNP with substitutional doping of one, two, and three silver atoms providing, therefore, $[Au_{24}Ag]^-$, $[Au_{23}Ag_2]^-$, $[Au_{22}Ag_3]^-$ compositions. All systems contain a single excess electron in line with the experimental evidence for $[Au_{25}(SR)_x]^-$, where $(SR)_x$ stands for the thiol-based ligands.[20] That is, the ground state of all systems is singlet.

Table 1. Simulated systems and simulation details. Proper equilibration of all systems was thoroughly controlled by analyzing evolution of many thermodynamic quantities, such as energy components, dipole moments, selected interatomic distances. Note that equilibration of the non-periodic systems occurs significantly faster due to absence of the long-order structure. Comments briefly explain starting system configurations

| # | # gold | # silver | # electrons | Time, ps | Comments |
|---|---|---|---|---|---|
| I | 24 | 1 | 1949 | 7×15+1×100 | identifying deepest minimum |
| II | 23 | 2 | 1911 | 4×50+1×100 | R (Ag-Ag) ≥ 0.8 nm |
| III | 23 | 2 | 1911 | 4×50+1×100 | $Ag_2$ at GNP surface |
| IV | 22 | 3 | 1879 | 4×70+1×100 | R (Ag-Ag) ≥ 0.6 nm |
| V | 22 | 3 | 1879 | 4×70+1×100 | $Ag_2$ & R ($Ag_2$-Ag) ≥ 0.6 nm |
| VI | 22 | 3 | 1879 | 4×70+1×100 | $Ag_3$ at GNP surface |

The PM7-MD simulation method obtains forces acting on every atomic nucleus from the electronic structure computation using the PM7 semiempirical Hamiltonian.[21] PM7 is a parameterized Hartree-Fock method, where certain integrals are pre-determined based on the well-known experimental data. Such a solution also allows for effective incorporation of the electron-correlation effects, while preserving a quantum-chemical foundation of the method.[21-24] Therefore, PM7 is able to capture any specific chemical interaction, such as hydrogen bonding, covalent bonding, metallic bonding, π-π stacking, etc.[21] PM7 is more physically relevant and mathematically sophisticated than any existing force field based technique. The derived forces are coupled with initial positions of atoms and randomly generated velocities (Maxwell-

Boltzmann distribution) at certain temperature. Subsequently, Newtonian equations-of-motions can be constructed and numerically integrated employing one of the available algorithms. This work relies on the velocity Verlet integration algorithm. Temperature may depend on good energy conservation and starting linear velocities given that the number of degrees of freedom is constant per system (constant energy ensemble). In turn, temperature may be adjusted periodically by rescaling atomic velocities in view of the reference (thermostat) temperature. This work employs a weak temperature coupling scheme[25] with a relaxation time of 0.15 ps, whereas the integration time-step equals to 2.5 fs. Different time-steps in the range of 0.5-5.0 fs were tested and the largest time-step satisfying total energy conservation criterion was chosen. Note that the simulated systems contain only heavy atoms. Therefore, a relatively large integration time-step is possible to accelerate trajectory sampling. Where applicable, the initial and final geometries of the doped GNPs were subject to geometry optimization using the improved eigenfollowing (EF) algorithm.[21]

More details of the PM7-MD implementation applied in the present study are described elsewhere.[16-19] The subsequent analysis was accomplished using formation energies, orbital energies, point partial charges, and pair correlation functions (PCFs). PCFs constitute an analogue of radial distribution functions for bulk matter, the only exception being normalization condition. The PCFs in this work were normalized so that the total area under each curve equals to unity. This solution allows comparing PCFs to one another to distinguish between different atom arrangements in the structures.

**Results and Discussion**

Seven PM7-MD annealing runs starting from arbitrary atomic configurations (system I) were performed to find the deepest possible minimum for the $[Au_{24}Ag]^-$ composition. First, the structure was uniformly cooled down from 2000 to 0 K during 15 ps. Second, the resulting cooled structure underwent the EF geometry optimization. Figure 1 summarizes relative deviations from the most stable identified configuration. Although the $[Au_{25}]^-$ GNP is extremely small, a series of local minima with very different depths was identified. The least energetically favorable configuration of $[Au_{24}Ag]^-$ is located ca. 23 percent of formation energy higher. The existence of such geometries in real world is unlikely, although they are predicted computationally. Even less favorable structures can be predicted if no annealing potential energy search precedes geometry optimization. Our previous work investigates this issue in great details.[16]

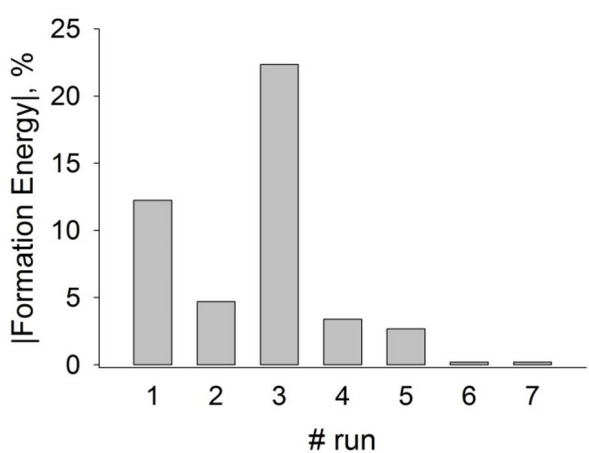

Figure 1. Relative formation energies of the $[Au_{24}Ag_1]^-$ GNPs obtained by the annealing PM7-MD simulations followed by the EF algorithm. See Table 1 for more methodology details.

According to Figure 2, significant fluctuations of the highest occupied molecular orbital (HOMO) and lowest unoccupied molecular orbital (LUMO) energy levels are observed as a function of formation energy (Figure 1). Fluctuations of HOMO are more significant than fluctuations of LUMO. Generally, smaller (more favorable) formation energy corresponds to smaller (more negative) energy of the HOMO level. Depending on the structure, the difference in HOMO can be as large as ca. 1 eV, which is very significant. The difference in LUMO, in turn, equals to ca. 0.7 eV. The discussed observations confirm a necessity to thoroughly look up correct GNP structures, which must be in concordance with the experimental results.

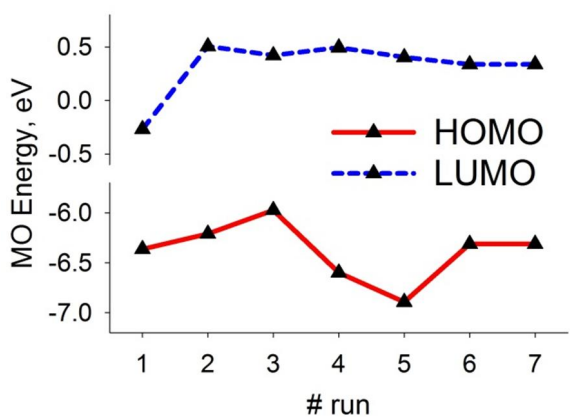

Figure 2. Energies of principal molecular orbitals, HOMO and LUMO, of the $[Au_{24}Ag_1]^-$ GNPs depending on the local minimum.

System II, $[Au_{23}Ag_2]^-$, features the following starting configuration. The two silver atoms were placed at the opposite sides of the doped GNP. The distance between them at the beginning of PM7-MD was at least 0.8 nm. Four independent PM7-MD simulations (50 ps long each) at 1500 K were performed to observe a real-time dynamics of the dopant fostered by energy gradient. All simulations yielded the $Ag_2$ clusters at the surface. The silver atoms approached one another quite quickly, within 15-35 ps in all cases. Consequently, the most energetically favorable configuration of $[Au_{23}Ag_2]^-$ (judged by energy of formation) was cooled down to 1000 K. The PCFs between gold atoms, gold-silver atoms, and silver-silver atoms were computed during 100 ps (Figure 3).

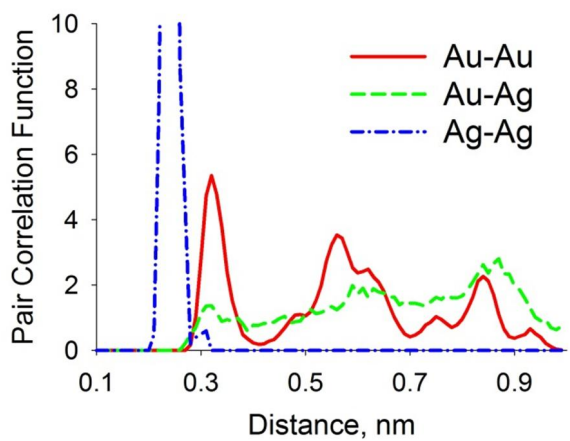

Figure 3. Pair correlation functions depicting a probability to find a given atom (Au, Ag) at certain distance from the reference atom (Au, Ag) in the $[Au_{23}Ag_2]^-$ GNP. The functions are based on the equilibrium 50 ps long PM7-MD trajectory at 1000 K maintained by thermostat.

The height of the first (and only) peak at the Ag-Ag PCF equals to 28 locating at 0.24 nm. Two is among the magic numbers for silver suggesting a superatom formation at the surface of GNP. A superatom is an arbitrary cluster of atoms that seem to exhibit some of the properties of elemental atoms. Two is interpreted as a number of electrons needed to fill the first electron shell. At the same time, binding of $Ag_2$ to the rest of $[Au_{23}Ag_2]^-$ appears very modest. Indeed, the first peak at Au-Ag PCF is located at 0.32 nm exhibiting a height of just 1.4. Compare with the height of the first peak at the Au-Au PCF, 5.4 and the same location, 0.32 nm. The Au-Au peaks are periodic indicating that the discussed GNP maintains a well-defined structure at 1000 K.

System III also features the $[Au_{23}Ag_2]^-$ composition, but the $Ag_2$ superatom was initially present in the initial configuration for PM7-MD. The four independent 50 ps long PM7-MD simulations were performed at 1500 K. None of the artificially constructed $Ag_2$ clusters decomposed until the end of the simulation. This observation is in perfect agreement with the behavior of system II. To recapitulate, $Ag_2$ clusters must be expected at the surface of GNPs, irrespective of their sizes.

Systems IV, V and VI contain the $[Au_{22}Ag_3]^-$ GNPs. System IV contains three Ag add-atoms at the GNP surface separated by at least 0.6 nm from the closest Ag add-atom. System V contains $Ag_2$ cluster, but the third Ag add-atom is separated by at least 0.6 nm. System VI contains $Ag_3$ cluster. Four instances of each of these systems were simulated during 70 ps at 1500 K. The most stable configuration was simulated for additional 100 ps at 1000 K to derive PCFs (Figure 4). Interestingly, all simulations indicated that the three silver add-atoms in $[Au_{22}Ag_3]^-$ exist as $Ag_2$ and Ag (all at the surface of GNP, Figure 5). The Ag-Ag first peak is smaller than in the case of $[Au_{23}Ag_2]^-$, 9 units (Figure 4). It is located at 0.27 nm as opposed to

0.24 nm in [Au$_{23}$Ag$_2$]$^-$. The Au-Ag coordination is better pronounced (2 units height) due to existence of the unpaired Ag. According to Figure 5, the unpaired Ag substitutes a gold atom to maintain the pyramid.

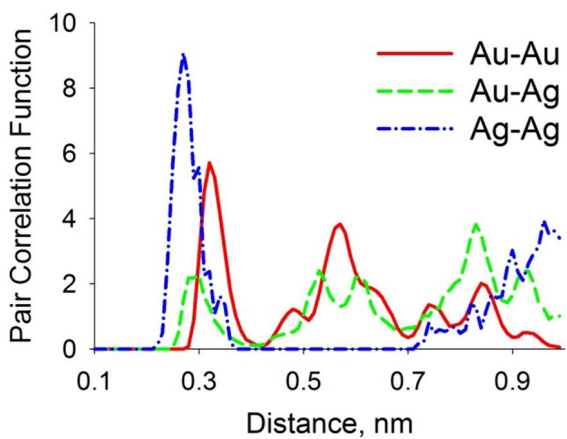

Figure 4. Pair correlation functions depicting a probability to find a given atom (Au, Ag) at certain distance from the reference atom (Au, Ag) in the [Au$_{22}$Ag$_3$]$^-$ GNP. The functions are based on the equilibrium 50 ps long PM7-MD trajectory at 1000 K maintained by thermostat.

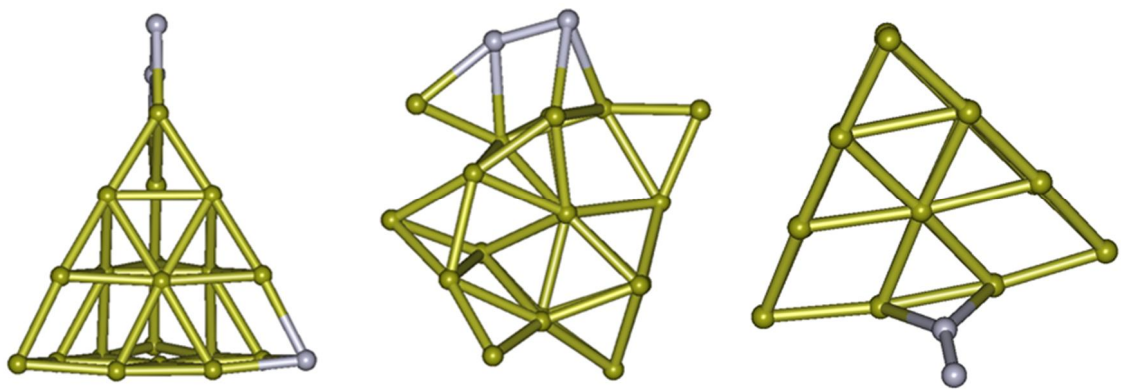

Figure 5. Structures of GNPs, [Au$_{22}$Ag$_3$]$^-$ (left) and [Au$_{23}$Ag$_2$]$^-$ (right). The configuration at the center is a less energetically favorable geometry of [Au$_{23}$Ag$_2$]$^-$. The optimized structures correspond to local minima on the potential energy surfaces obtained using the annealing PM7-MD simulations supplemented by the EF geometry optimization algorithm.

Distribution of point electrostatic charges (Figure 6) provides an integral measure of electron density anisotropy with the nanostructure. Recall that all considered GNPs are negatively charged. An excess electron is used to fill all shells and avoid a doublet electronic state. The observed charge distribution indicated that only a few atoms (three in the case of [Au$_{22}$Ag$_3$]$^-$ and four in the case of [Au$_{23}$Ag$_2$]$^-$) are responsible to accommodate an excess electron, while all other gold atoms remain nearly neutral. An increased chemical reactivity may be expected at these charged Au atoms. For this reason, coverage of GNPs by the thiol-based groups is absolutely necessary.

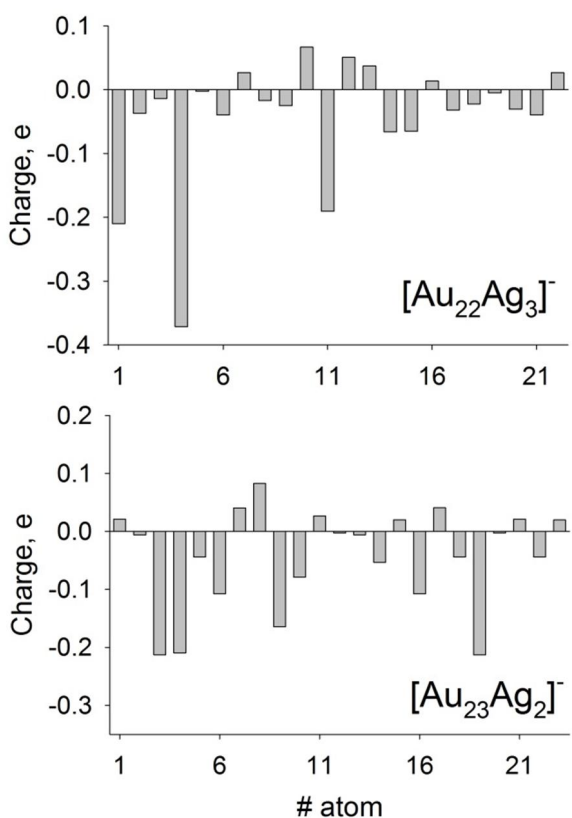

Figure 6. Point electrostatic charges on the Au atoms constituting the two studied GNPs, [Au$_{22}$Ag$_3$]$^-$ and [Au$_{23}$Ag$_2$]$^-$. The charges were calculated according to the well-established Coulson's charge assignment scheme based on the population analysis. The depicted numbers of atoms correspond to consecutive numbering of Au atoms in the initial GNP geometry. It does not reflect any chemical bonding occurring as the result of annealing PM7-MD and EF geometry optimization.

Another point of interest is fractional charges at Ag add-atoms. In [Au$_{24}$Ag]$^-$, silver appears slightly positive, +0.09e. It is in concordance with electronegativity of elements, 2.54 (Au) and 1.93 (Ag). A superatom is formed in [Au$_{23}$Ag$_2$]$^-$, therefore, we expect certain fraction of covalent bonding between the involved Ag add-atoms, which would alternate charges. Indeed, the charges exhibit opposite signs, -0.08e and +0.06e. Similar pattern is also observed in $[Au_{22}Ag_3]^-$, while the unpaired Ag atom is nearly neutral, -0.01e.

**Conclusions**

The PM7-MD simulations on six systems featuring silver-doped gold nanoparticles, $[Au_{25-N}Ag_N]^-$, were conducted. Multiple simulations with different starting points on the phase space were used to locate reliable structures. Formation energies, orbital energy levels, distribution of electron density, and pair correlation functions were derived to characterize the obtained silver-doped GNPs. It was found that all dopant atoms prefer to locate at the surface of the nanoparticle. If more than a single Ag add-atom is present, $Ag_2$ cluster, which can be called a superatom by definition, always forms. Remarkably, the assembly of a superatom is extremely fast, 15-35 ps at 1500 K. Therefore, this process must be highly expected during the formation of GNPs and their doping. Note that dynamics of atoms at the surface is always much faster than their dynamics in bulk material. If more than two Ag add-atoms are present, they still form $Ag_2$ cluster, whereas the remaining atom substitutes an original gold atom to maintain an overall geometry of GNP. The $Ag_3$ cluster is not favored, irrespective of the starting configuration.

The reported results are important for projecting doped nanostructures. Determination of GNP geometries is necessary, since electronic structure of the nanoscale particle and, therefore, a large set of its properties, strongly depend on the internal atom arrangement. The knowingly performed theoretical simulations provide key physical insights many of which are inaccessible for current experimental techniques.

**Acknowledgments**

This investigation has been partially supported by the research grant from CAPES under the "Science Without Borders" program.

# REFERENCES

(1) Qian, H. F.; Zhu, M. Z.; Wu, Z. K.; Jin, R. C. Quantum Sized Gold Nanoclusters with Atomic Precision. *Accounts of Chemical Research* 2012, *45*, 1470-1479.
(2) Jiang, D. E. The Expanding Universe of Thiolated Gold Nanoclusters and Beyond. *Nanoscale* 2013, *5*, 7149-7160.
(3) Baia, M.; Baia, L.; Astilean, S. Gold Nanostructured Films Deposited on Polystyrene Colloidal Crystal Templates for Surface-Enhanced Raman Spectroscopy. *Chemical Physics Letters* 2005, *404*, 3-8.
(4) Beqa, L.; Singh, A. K.; Fan, Z.; Senapati, D.; Ray, P. C. Chemically Attached Gold Nanoparticle-Carbon Nanotube Hybrids for Highly Sensitive Sers Substrate. *Chemical Physics Letters* 2011, *512*, 237-242.
(5) Bera, M. K.; Sanyal, M. K.; Banerjee, R.; Kalyanikutty, K. P.; Rao, C. N. R. Effect of Vibrations on the Formation of Gold Nanoparticle Aggregates at the Toluene-Water Interface. *Chemical Physics Letters* 2008, *461*, 97-101.
(6) Dasary, S. S. R.; Rai, U. S.; Yu, H. T.; Anjaneyulu, Y.; Dubey, M.; Ray, P. C. Gold Nanoparticle Based Surface Enhanced Fluorescence for Detection of Organophosphorus Agents. *Chemical Physics Letters* 2008, *460*, 187-190.
(7) Fajin, J. L. C.; Cordeiro, M. N. D. S.; Gomes, J. R. B. Dft Study on the No Oxidation on a Flat Gold Surface Model. *Chemical Physics Letters* 2011, *503*, 129-133.
(8) Grant, C. D.; Schwartzberg, A. M.; Yang, Y. Y.; Chen, S. W.; Zhang, J. Z. Ultrafast Study of Electronic Relaxation Dynamics in Au-11 Nanoclusters. *Chemical Physics Letters* 2004, *383*, 31-34.
(9) Kasture, M.; Sastry, M.; Prasad, B. L. V. Halide Ion Controlled Shape Dependent Gold Nanoparticle Synthesis with Tryptophan as Reducing Agent: Enhanced Fluorescent Properties and White Light Emission. *Chemical Physics Letters* 2010, *484*, 271-275.
(10) Kim, K. K.; Kwon, H. J.; Shin, S. K.; Song, J. K.; Park, S. M. Stability of Uncapped Gold Nanoparticles Produced by Laser Ablation in Deionized Water: The Effect of Post-Irradiation. *Chemical Physics Letters* 2013, *588*, 167-173.
(11) Yamamuro, S.; Sumiyama, K. Why Do Cubic Nanoparticles Favor a Square Array? Mechanism of Shape-Dependent Arrangement in Nanocube Self-Assemblies. *Chemical Physics Letters* 2006, *418*, 166-169.
(12) Zheng, B.; Uenuma, M.; Okamoto, N.; Honda, R.; Ishikawa, Y.; Uraoka, Y.; Yamashita, I. Construction of Au Nanoparticle/Ferritin Satellite Nanostructure. *Chemical Physics Letters* 2012, *547*, 52-57.
(13) Rosi, N. L.; Mirkin, C. A. Nanostructures in Biodiagnostics. *Chemical Reviews* 2005, *105*, 1547-1562.
(14) Garg, N.; Mohanty, A.; Lazarus, N.; Schultz, L.; Rozzi, T. R.; Santhanam, S.; Weiss, L.; Snyder, J. L.; Fedder, G. K.; Jin, R. C. Robust Gold Nanoparticles Stabilized by Trithiol for Application in Chemiresistive Sensors. *Nanotechnology* 2010, *21*.
(15) Li, G.; Jin, R. C. Atomically Precise Gold Nanoclusters as New Model Catalysts. *Accounts of Chemical Research* 2013, *46*, 1749-1758.
(16) Andreeva, N. A.; Chaban, V. V. Global Minimum Search Via Annealing: Nanoscale Gold Clusters. *Chemical Physics Letters* 2015, *622*, 75-79.
(17) Chaban, V. Competitive Solvation of (Bis)(Trifluoromethanesulfonyl)Imide Anion by Acetonitrile and Water. *Chemical Physics Letters* 2014, *613*, 90-94.
(18) Chaban, V. Annealing Relaxation of Ultrasmall Gold Nanostructures. *Chemical Physics Letters* 2015, *618*, 46-50.
(19) Chaban, V. The Thiocyanate Anion Is a Primary Driver of Carbon Dioxide Capture by Ionic Liquids. *Chemical Physics Letters* 2015, *618*, 89-93.

(20)     Heaven, M. W.; Dass, A.; White, P. S.; Holt, K. M.; Murray, R. W. Crystal Structure of the Gold Nanoparticle [N(C8h17)(4)][Au-25(Sch2ch2ph)(18)]. *Journal of the American Chemical Society* 2008, *130*, 3754-3755.
(21)     Stewart, J. J. P. Optimization of Parameters for Semiempirical Methods Vi: More Modifications to the Nddo Approximations and Re-Optimization of Parameters. *Journal of Molecular Modeling* 2013, *19*, 1-32.
(22)     Stewart, J. J. P. Optimization of Parameters for Semiempirical Methods V: Modification of Nddo Approximations and Application to 70 Elements. *Journal of Molecular Modeling* 2007, *13*, 1173-1213.
(23)     Stewart, J. J. P. Application of the Pm6 Method to Modeling the Solid State. *Journal of Molecular Modeling* 2008, *14*, 499-535.
(24)     Stewart, J. J. P. Application of the Pm6 Method to Modeling Proteins. *Journal of Molecular Modeling* 2009, *15*, 765-805.
(25)     Berendsen, H. J. C.; Postma, J. P. M.; Vangunsteren, W. F.; Dinola, A.; Haak, J. R. Molecular-Dynamics with Coupling to an External Bath. *Journal of Chemical Physics* 1984, *81*, 3684-3690.